\documentstyle[twoside,fleqn,espcrc2]{article}

\title{Multiple mass solvers}

\author{B. Jegerlehner\address{Indiana University, Depatrment of Physics,
	                       Swain Hall West 117,\\Bloomington, IN 47408, USA}}

\begin{document}

\begin{abstract}
We present a general method to construct multiple mass solvers from standard 
algorithms. As an example, the BiCGstab-M algorithm is derived.
\end{abstract}

\maketitle

\section{INTRODUCTION}

It has been discussed recently that, using Krylov space solvers, 
the solutions of shifted linear equations, 
where $(A+\sigma) x - b = 0$ 
has to be calculated for a whole set of values of $\sigma$,
can be found at the cost of only one inversion.
This kind of
problem arises in quark propagator calculations for QCD as well as
other parts of computational physics (see \cite{qmr}).
It has been realized that several algorithms allow to perform this task 
using only
as many matrix-vector operations as the solution of the most 
difficult single system requires. This has been achieved for the 
QMR \cite{qmr}, the MR \cite{m3r} and the Lanczos-implementation
of the BiCG method \cite{borici}. We present here a unifying discussion
of the principles to construct such algorithms and succeed in constructing
shifted versions of the CG, CR, BiCG and BiCGstab algorithms \cite{myself}, using only two 
additional vectors for each mass value.

\section{METHOD}

The iterates of Krylov space methods, especially the residuals $r_n$,
are generally polynomials $P_n(A)$ of the matrix applied to some initial 
vector $r_0$. The polynomial generating the residual generally has to be an
approximation to zero in some region enclosing the spectrum of $A$ while
statisfying $P_n(0) = 1$.
The key to the method is the observation that
shifted polynomials, defined by
\begin{equation}\label{eq1}
P^\sigma_n(A+\sigma) = c_n^\sigma P_n(A) ,
\end{equation}
are useful objects, since vectors generated by these shifted
polynomials can be calculated for multiple $\sigma$ values using no additional
matrix-vector multiplications.  
We expect $c_n^\sigma < 1$ if the condition number of $A+\sigma$ is smaller than
the one of $A$, which is confirmed in numerical tests. 

Generally the polynomial generated in a solver is defined by some recursion 
relation. We will therefore need to know the recursion relation for the
shifted polynomial, too, which can be found easily by parameter matching.
Here, we discuss only polynomials which statisfy $P(0)=1$, but
more general normalisation conditions are handled analogously.
For the two-term recursion relation 
\begin{equation}
P_{n+1}(x) = (\alpha_n x + 1) P_n(x)
\end{equation}
we find for the polynomial shifted by $\sigma$
\begin{eqnarray}
P_{n+1}^\sigma(x) &=& \left({\alpha_n \over 1-\sigma\alpha_n} x + 1\right) P^\sigma_n(x) \label{line1}\\
                  &=& \left(\prod_i {1\over 1-\sigma\alpha_i}\right) P_{n+1}(x-\sigma) \label{line2}.
\end{eqnarray}
This formula has also been found in \cite{m3r} with different methods. From
(\ref{line1}) we can read off the parameters of the shifted polynomial, while 
(\ref{line2}) determines $c_n^\sigma$.
Note that this formula holds for any choice of $\alpha_n$. We can easily
generalize this method to three-term recurrences 
\begin{equation}
P_{n+1}(x) = (\alpha_n x + \beta_n) P_n(x) + \gamma_n P_{n-1}(x)
\end{equation}
and find explicit expressions for the parameters of the shifted polynomial.
It turns out that the Lanczos polynomials for the matrices $A$ and $A+\sigma$
fulfill (\ref{eq1}) automatically 
(this was the original observation in \cite{qmr}). We derived the shifted
polynomial however for arbitrary choices of parameters $\alpha$ and $\beta$.
The most interesting case are coupled two-term recurrences, since they
have superior stability properties over three-term recurrences.
We consider the general recurrence
\begin{eqnarray}
Q_n(x) &=& P_n(x) + \alpha_n Q_{n-1}(x) \\
P_{n+1}(x) &=& P_n(x) + \beta_n x Q_n(x) .
\end{eqnarray}
In $CG$-type algorithms, the parameters are chosen so that $P_n(x)$ is the
Lanzcos-polynomial (normalized to $P(0)=1$). We thus demand $P^\sigma_n(x+\sigma) = c_n^\sigma P_n(x)$. By transforming the above 
relation to a simple three-term recurrence and applying the formulae found in this case for the shifted parameters we find 
\begin{eqnarray}
c_{n+1}^\sigma &=& {c_n^\sigma \over
{\beta_n\over \beta_{n-1}} \alpha_n \left(1- {c_n^\sigma\over c_{n-1}^\sigma}\right) +  
(1 - \sigma \beta_n)} \\
\beta_n^\sigma &=& \beta_n {c_{n+1}^\sigma \over c_n^\sigma} \\
\alpha_n^\sigma &=& \alpha_n {c_n^\sigma \beta_{n-1}^\sigma \over
c_{n-1}^\sigma \beta_{n-1}}  .
\end{eqnarray}
It can easily be checked that $Q_n^\sigma(x+\sigma) \ne c Q_n(x)$, so if we
want to use this recursion relation in an algorithm we have to replace
\begin{equation}
\beta_n^\sigma (x+\sigma) Q_n^\sigma(x) = c_{n+1}^\sigma P_{n+1}(x) - c_n^\sigma P_n(x)
\end{equation}
in the shifted systems. 
Since in CG-type algorithms the update of the solution vector
involves $Q_n(x) v_0$, this vector has to be iterated and stored for all
shifted systems.

\section{APPLICATIONS}

Using the above formulae, we can easily derive a variety of linear system
solvers as shown in table \ref{tab1}. 
\begin{table}
\begin{center}
\begin{tabular}{|c|c|c|c|c|}
\hline
Method & Reference & Memory \\
\hline
MR-M & \cite{m3r,myself} & $N$ \\
CR-M & \cite{myself} & $2N$ \\
QMR-M (3-term) & \cite{qmr} & $3N$ \\
QMR-M (2-term) & \cite{qmr2,myself} & $3N$ \\
TFQMR-M & \cite{qmr} & $5N$ \\
BIORESU & \cite{borici,myself} & $2N$ \\
BiCG-M & \cite{myself} & $2N$ \\
BiCGstab-M & \cite{myself} & $2N+1$ \\
\hline
\end{tabular}
\end{center}
\caption{\label{tab1}Memory requirements and references for shifted
system algorithms for unsymmetric or nonhermitean matrices. We list the
number of additional vectors neccessary for $N$ additional values of $\sigma$ (
which is independent of the use of the $\gamma_5$-symmetry). }
\end{table}
We present here only the algorithm of greatest interest for quark propagator
calculations, BiCGstab-M.

\subsection{BiCGstab-M}

The BiCGstab-M algorithm is a mixture between the BiCG and the MR algorithm.
It is therefore not surprising that we can simply use the formulae for
the two-term and the coupled two-term recurrences and construct a shifted
algorithm. 
In the BiCGstab algorithm \cite{bicgstab}, we generate the following sequences
\begin{eqnarray}
r_n &=& P_n(A) R_n(A) r_0 \\
w_n &=& P_n(A) R_{n-1}(A) r_0 \\
s_n &=& Q_n(A) R_n(A) r_0
\end{eqnarray}
where $Z_n(x)$ and $Q_n(x)$ are exactly the 
BiCG-polynomials and $R_n(x)$ is derived
from a minimal residual condition.
For the shifted algorithm we demand
\begin{equation}\label{shiftc}
P_n^\sigma(x+\sigma) = c_n^\sigma P_n(x) ,~
R_n^\sigma(x+\sigma) = d_n^\sigma R_n(x) .
\end{equation}
Using the above formulae we can explicitly determine the constants $c$ and $d$
and the shifted parameters ot the polynomials. The remaining difficulty is
to derive the iteration for the solution $x_n$ and the vector $s_n$.
The update of these two vectors has the form
\begin{eqnarray}
x_{n+1} &=& x_n - \beta_n s_n + \chi_n w_{n+1} \\
s_{n+1} &=& r_{n+1} + \alpha_{n+1}(s_n - \chi_n A s_n)
\end{eqnarray}
This means we have to eliminate $A s_n$ from the update of $s_n$.
The updates for the shifted vectors
$x_n^\sigma$ and $s_n^\sigma$ then look as follows:
\begin{eqnarray}
x_{n+1}^\sigma &=& x_n^\sigma - \beta_n^\sigma s_n^\sigma + \chi_n^\sigma c_n^\sigma d_{n-1}^\sigma w_{n+1} \\
s_{n+1}^\sigma &=& c_{n+1}^\sigma d_{n+1}^\sigma r_{n+1} + \alpha_{n+1}^\sigma
\times \\
&& \left(
s_n^\sigma - {\chi_n^\sigma \over \beta_n^\sigma}(c_{n+1}^\sigma d_n^\sigma w_{n+1} -
c_n^\sigma d_n^\sigma r_n)\right) \nonumber
\end{eqnarray}
We therefore need to introduce 2 vectors for each shifted system and one
additional vector to store $r_n$. Note that the case $\beta_n = 0$ leads to a breakdown of the BiCGstab algorithm
and does not introduce any new problems for the shifted method.

The convergence of the shifted algorithms 
can be verified by checking that $c_n r_n \le 1$ .
It is however generally advisable for all shifted algorithms to test all 
systems for convergence after
the algorithm finishes since a loss of the conditions (\ref{shiftc}) due
to roundoff errors might lead to erratic convergence. 

\section{LIMITATIONS}

The most serious limitation of the method is given by the fact that the
starting residual for all systems must be the same, which excludes 
$\sigma$-dependent left preconditioning. Furthermore, preconditioning must
retain the shifted structure of the matrix. This means that especially
even-odd preconditioning is not applicable. To stabilize the algorithm,
however, one can apply polynomial preconditioning:
\begin{equation}
P(A) A y = b, \quad x = P(A) y .
\end{equation}
We must have
\begin{equation}\label{condpre}
P^\sigma(A+\sigma) (A+\sigma) = P(A) A + \eta .
\end{equation}
Note that $P^\sigma$ might not be a good preconditioner for $A+\sigma$, 
which is compensated for by the faster convergence of the shifted system.
A linear preconditioner, which was also proposed in \cite{m3r}, is given by
\begin{equation}
P^\sigma(x) = 2(\sigma + m) - x .
\end{equation}
For the Wilson and clover matrix, this polynomial has the property that $P^\sigma(x)$ is a good preconditioner
for $A+\sigma$. This preconditioner has been found to work well in those cases.
Higher order polynomials can be derived from condition (\ref{condpre}). 

\section{CONCLUSIONS}

We presented a simple point of view to understand the structure of
Krylov space algorithms for shifted systems, allowing us to construct
shifted versions of most short recurrence Krylov space algorithms.
The shifted CG-M and CR-M algorithm can be applied
to staggered fermion calculations. Since efficient 
preconditioners for the staggered fermion matrix are not known, 
a very large improvement by these algorithms can be expected.
We presented the BiCGstab-M 
method, which, among the shifted algorithms, 
is the method of choice for quark propagator calculations
using Wilson (and presumably also clover) fermions if enough memory 
is available. 
The numerical stability of the algorithms has been found to be good \cite{myself}.
Roundoff errors might however in some cases affect the convergence of the
shifted systems so that the final residuals have to be checked. 
Other discussions can be found in \cite{qmrfromm,compmulti}.

\section*{Acknowledgements}

This work was supported in part by the U.S. Department
of Energy under Grant No. DE-FG02-91ER40661. 
I would like to thank S. Pickels, C. McNeile and S. Gottlieb for
helpful discussions.


\begin{thebibliography}{9}
 
\bibitem{qmr} R. W. Freund, Solution of shifted linear systems by quasi-minimal
residual iterations, Numerical Linear Algebra, L. Reichel, A. Ruttan and
R.S. Varga (eds.), Berlin: W. de Gruyter 1993, 101--121.

\bibitem{m3r} U. Gl\"assner, S. G\"usken, T. Lippert, G. Ritzenh\"ofer,
K. Schilling and A. Frommer, 
hep-lat/9605008

\bibitem{borici} A. Bori\c{c}i, SCSC report
TR-96-27

\bibitem{myself} B. Jegerlehner, IUHET-353, hep-lat/9612014

\bibitem{bicgstab} H.A. van der Vorst, SIAM J. Sci Statist. Comput. 13 (1992) 631,
M. H. Gutknecht, 
SIAM J. Sci. Comput. 14 (1993) 1020



\bibitem{qmr2} R.W. Freund and N.M. Nachtigal, An implementation of the {QMR} method based on coupled two-term recurrences, http://cm.bell-labs.com/cm/cs/doc/92/4-06.ps.gz

\bibitem{qmrfromm} A. Frommer, B. N\"ockel, S. G\"usken, T. Lippert and K. Schilling, 
Int. J. Mod. Phys C6 (1995) 627

\bibitem{compmulti} H-P. Ying, S-J. Dong and K-F. Liu,  
Nucl. Phys. Proc. Suppl 53 (1997) 993

\end{thebibliography}
\end{document}